# One-Dimensional Potassium Chains on Silicon Nanoribbons

Tongtong Chen[#], Wenjia Zhang[#], Xiaobei Wan , Xiaohan Zhang, Yashi Yin, Jinghao Qin, Fengxian Ma *, Juntao Song *, Ying Liu, and Wen-Xiao Wang *

**ABSTRACT**

Silicon nanoribbons (SiNRs), characterized by a pentagonal structure composed of silicon atoms, host one-dimensional (1D) Dirac Fermions and serve as a minimalist atomic template for adsorbing various heteroatoms. Alkali-metal (AM) atoms, such as Na and K, with electronic structures comparable to those of hydrogen are of particular interest for such adsorption studies. However, the adsorption of AM atoms on SiNRs and its tunation on the properties of SiNRs have not yet been fully explored. In this study, we examined the adsorption of K atoms on high-aspect-ratio SiNRs and the resultant electronic properties using a combination of scanning tunneling microscopy (STM) and density functional theory calculations. K atoms prefer to adsorb on double- and multi-stranded SiNRs owing to the low adsorption energies at these sites. Each K atom and its three nearest Si atoms exhibit a triangular morphology resulting from charge transfer between K and Si atoms, as verified by theoretical calculations. As the K coverage of the SiNRs increased, the K atoms organize into 1D zigzag chains on the SiNRs. Moreover, K adsorption on the SiNRs was determined to be reversible. The deposition of K atoms on the SiNRs was achieved using a voltage pulse of the STM tip, without damaging the SiNRs structure. In addition, K adsorption effectively modulates the Dirac cone position of the SiNRs relative to the Fermi level. This study unveils the adsorption mechanism of AM atoms on SiNRs, providing a useful approach for heteroatom adsorption on other nanoribbons.

## 1. INTRODUCTION

As the first confirmed two-dimensional Dirac material, graphene exhibits a series of extraordinary physical phenomena due to its unique band structure, including the extremely high carrier mobility[1, 2], quantum (spin) Hall effect[3-6], correlated physics[7, 8] and so on. Dirac materials are a special class of crystals whose electronic bands form linear intersections near the Fermi level, creating Dirac cones and exhibiting distinctive zero-gap characteristics. Following graphene, other 2D materials like silicene[9-14], germanene[15-17], and borophene[18] have been theoretically predicted and experimentally confirmed as Dirac materials using STM and ARPES techniques. Recently, researchers have turned their attention to one-dimensional (1D) materials. Silicon nanoribbons (SiNRs) are one of them, containing 1D massless Dirac fermions, and experimentally, the existence of Dirac cones has been confirmed through ARPES[19]. Moreover, SiNRs also exhibit strong resistance to oxidation[20]. As an important component of low-dimensional Dirac materials, SiNRs are propelling the development of miniaturized electronic device technology.

Alkali metals such as Li and K, with their hydrogen-like simple electronic structures, serve as ideal reference systems for studying atomic adsorption phenomena. While the adsorption of alkali metals on two-dimensional Dirac materials (e.g., graphene and silicene) has been extensively investigated[21-29]. For example, the K atoms on silicene form an ordered close-packed silicene$(3 \times 3)$ phase, and the "V"-shaped density of states is considered to be the Dirac feature of silicene. On the other hand, the interactions between AMs and the newly discovered one-dimensional Dirac material—SiNRs—remain largely unexplored. This knowledge gap is striking given the demonstrated significance of SiNR-based adsorption systems. For instance, studies have shown that transition metals like Mn and Co adsorbed on SiNRs exhibit ferromagnetic properties under specific conditions[30-33] and the selective adsorption of Mn atoms with configurational adaptability can form Mn silicide structures[30], highlighting their potential for spintronic applications.

However, neither macroscopic transport studies nor atomic-scale experimental investigations have systematically addressed the adsorption of alkali metals on SiNRs. Such research is critically needed to unravel how the unique SiNRs modulate charge transfer, electronic doping, and interfacial interactions with alkali metals—factors that are pivotal for designing next-generation low-dimensional energy storage systems.

In this study, we investigated the adsorption behavior of K atoms on high-aspect-ratio 1D SiNRs and the resultant electronic properties using scanning tunneling microscopy (STM) and STS. By comparing the deposition of K atoms onto the SiNRs at room temperature (293K) and low-temperature (150 K), we determined that K provides a higher degree of coverage at the lower temperature. Furthermore, regardless of the temperature (room or low temperature), K atoms preferentially adsorb to sites between the individual strands of double-stranded NRs (DNRs) or triple-stranded NRs (TNRs). Each K atom on the SiNRs forms a "triangular" structure with Si because of charge transfer between each K atom and its three nearest Si neighbors. Additionally, as the K coverage of the SiNRs increased, the K atoms form a 1D zigzag chain-like structure along the SiNRs. By applying *in situ* tip pulses, we could change the K-atom adsorption sites and control the transition between two K adsorption orientations. We also found that the K atoms can be desorbed by applying a large pulse, without damaging the SiNRs. This feature suggests that K atom adsorption onto the SiNRs constitutes a reversible physical adsorption process. In terms of electronic properties, the STS of the SiNRs exhibit a "dip" feature below the Fermi level, which corresponds to the Dirac point in the band structure of the SiNRs. This is because as the coverage of K atoms increases, charge transfer occurs between the K atoms and SiNRs, causing the Dirac point to shift below the Fermi level. The entire process is similar to electronic doping and indicates ionic interactions between the two materials. Our work not only adds new knowledge to the field of AM atom adsorption on 1D Dirac materials but also serves as an important reference for heteroatom adsorption on 1D NRs.

## 2. METHODS

The K adsorption experiments were conducted in a combined system of ultrahigh-vacuum low-temperature STM and molecular beam epitaxy from CreaTec, using a base vacuum of 7×10$^{-11}$ mbar. First, SiNRs were epitaxially grown on single-crystal Ag(110) substrates. The surface of the single-crystal Ag(110) substrate was cleaned in situ via repeated cycles of Ar+ sputtering, followed by annealing at approximately 963 K. Silicon was deposited on the cleaned Ag(110) substrate held at 473 K by heating a silicon wafer (3 mm×10 mm) using a simple custom-built silicon wafer evaporation device with direct-current power supply, and the SiNRs sample was obtained under the same conditions for 16 min.

Once the sample naturally cools to room temperature, the deposition of K at room temperature can be carried out. For low-temperature K deposition, the SiNRs were cooled to approximately 150 K (150 ± 1 K) with a continuous flow of liquid nitrogen. After K deposition, the sample was immediately transferred to the STM scanning chamber for measurement using an etched tungsten tip at 77 K. The bias voltage is defined as the sample bias with respect to the tip. The STM images were processed using WSxM. STS measurements were accomplished using a lock-in amplification technique, which involved the implementation of a small sinusoidal modulation (676 Hz, 10 mV) of the sample bias voltage using an integrated lock-in amplifier module in the Nanonis controller. The STS spectra were calibrated using a clean Ag(111) surface.

## 3. RESULTS AND DISCUSSION

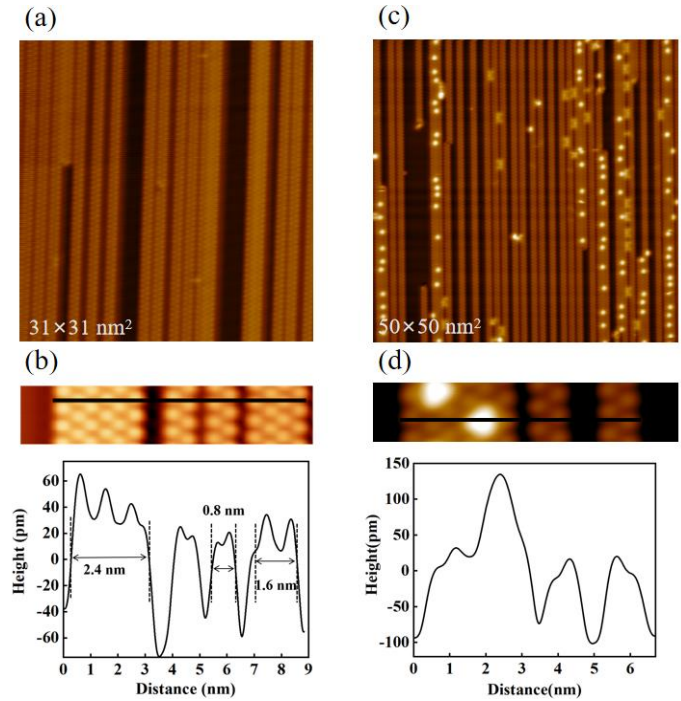

**Figure 1.** Typical STM topographies of SiNR/Ag(110) before and after K adsorption. (a) STM topography of the SiNRs grown on Ag(110) (31×31 nm$^2$, $V$ = 800 mV, $I$ = 50 pA). (b) Atomic-resolution image of the single-, double-, and triple-stranded SiNRs, along with a line profile

corresponding to the black solid line drawn in (b). (c) STM topography of SiNRs/Ag(110) after the room-temperature deposition of an appropriate amount of K adsorbate atoms (50×50 nm$^2$, $V = -800$ mV, $I = 100$ pA).

To study the K adsorption behavior of 1D SiNRs with different widths, the SiNRs were grown at a substrate temperature of 473 K, which is conducive to the formation of single- to multi-stranded SiNRs[34-36]. The Si atoms were deposited on Ag(110) at a substrate temperature of 473 K for 16 min to form SiNRs of different widths. Figure 1a shows the STM topography images of the 1D SiNRs. The z profile corresponding to the black line in Figure 1b reveals that the SNR (width = 0.8 nm) has ×3 periodicity along the [001] direction on Ag(110). The DNR, which has a width of 1.6 nm, displays ×5 periodicity, while the TNR, which has a width of 2.4 nm, exhibits ×7 periodicity, both along the [001] direction, as shown in Figure 1b. These features of the SiNRs agree well with those reported previously.[19, 34]

A direct current power supply is used to regulate the current at the K source to 5.2 A and the voltage to 2.5 V, releasing K onto the surface of an Ag(110) substrate at a temperature of 293 K. After the deposition of a certain number of K atoms on the SiNRs, bright spots corresponding to the adsorbed K atoms become apparent in the STM image, as shown in Figure 1c. The K atoms locate approximately 100-120 pm above the SiNRs, as shown in the lower panel of Figure 1d. Surprisingly, the distribution of K atoms is not uniform. The majority of the deposited K atoms exhibit a preference for DNRs and TNRs, with relatively fewer K atoms adsorbing to SNRs. That is, the K atoms exhibit pronounced selectivity in adsorbing to specific SiNR sites on the substrate. Moreover, as the K coverage increased, the adsorbed K atoms tend to form an ordered chain-like pattern on the DNRs and TNRs. These observations prompted further exploration of the specific adsorption patterns of the K atoms on the SiNRs.

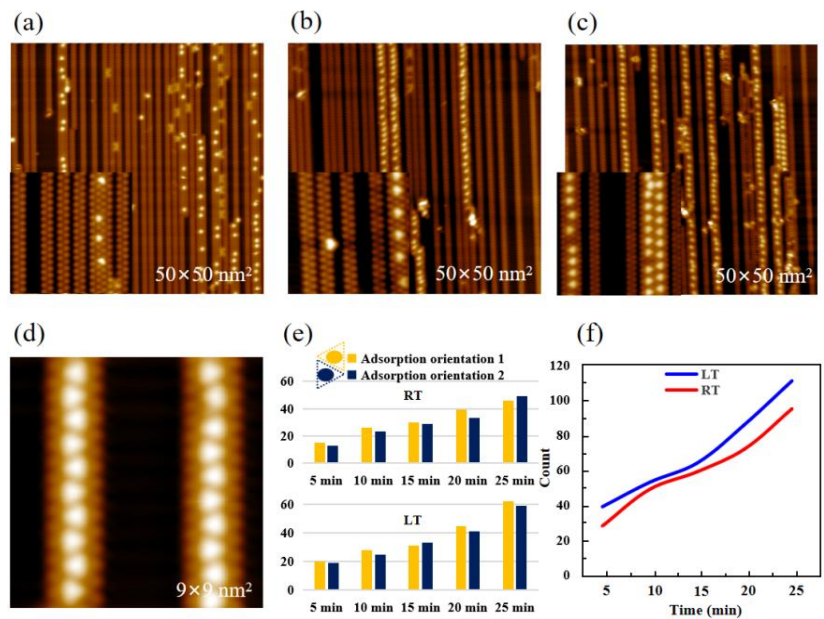

**Figure 2.** (a)-(c) STM topographical and atomic-resolution images obtained after the deposition of K atoms on SiNRs at room temperature for different durations: (a) 5 min (50 ×50 nm$^2$, $I = 50$ pA,

$V$ = 800 mV; The inset image is 10 ×10 nm$^2$, $I$ = 80 pA, V = −1 V), (b) 15 min (50 ×50 nm$^2$, $I$ = 50 pA, $V$ = −1 V; the inset image is 10 ×10 nm$^2$, $I$ = −80 pA, $V$ = −1 V and (c) 25 min (50 ×50 nm$^2$, $I$ = 50 pA, $V$= −1 V; the inset image is 12 ×12 nm$^2$, $I$ = 80 pA, V = −800 mV). (d) STM topography image of K atoms adsorbed to 1D ordered "Zigzag". (e) Statistical graphs of the two adsorption orientations of K atoms as a function of the adsorption time at 150 K and room temperature (293 K), respectively. The measured temperature by STM is 77K. (f) Total number of K atoms for low-temperature (150 K) and room-temperature adsorption.

Figures 2a-2c shows the typical STM topography images obtained after the adsorption of K atoms on the SiNRs at room temperature for 5, 15, and 25 min. As the deposition time increased, the coverage of K increased, and a transition from discrete K atoms to a chain-like structure was observed. At different coverage levels, only a small number of K atoms tend to adsorb at the edges of silicon nanoribbons, as depicted in the inset of Figure 2b. The majority of K atoms preferentially adsorb between the individual strands of DNRs, further underscoring the site-selective adsorption behavior of K atoms on the SiNRs. We also investigated the adsorption of K atoms at room temperature; a lesser number of K atoms adsorb at the SiNRs compared with that at 150K after the same adsorption time, as shown in Figure 2f. This phenomenon is reasonable, because the diffusion of K atoms is faster at room temperature than at 150 K. Notably, all K atoms exhibit a triangular structure with two orientations in the middle of the DNRs. By statistically analyzing the two K adsorption orientations on the DNRs and TNRs, we determine that the occurrence of these orientations tend to equalize as deposition time increasing, as shown in Figure 2e. This trend was consistently observed under both low- and room-temperature conditions. These results indicate that the small energy difference between the two orientations of triangular adsorption sites plays a significant role in this phenomenon. Additionally, we observed the morphology of K atoms adsorbed to a near-saturated state, as shown in Figure 2d, which provides further assistance in our study of the adsorption sites of K atoms.

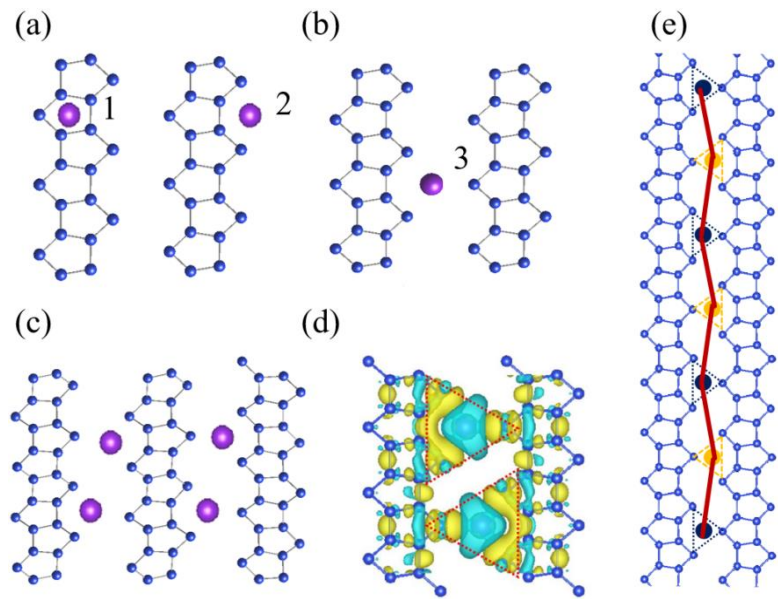

**Figure 3.** (a)-(c) Computational structural models for the adsorption of K atoms on the SNR, DNR, and TNR; the purple atoms represent K and the blue atoms represent Si. (d) Top view of the charge density difference map for the K/SiNR system, where the red dashed triangles correspond to the morphological characteristics after K atom adsorption, and the green and yellow areas represent the electron-deficient and electron-enriched areas, respectively. (e) Schematic of the one-dimensional ordered "zigzag" pattern of K atoms formed on SiNRs.

**Table 1**

Adsorption energies of K atoms at different sites on SiNRs

| Structure | $E_{ads}$ (eV) |
|---|---|
| K/SiNR-1 | −3.40 |
| K/SiNR-2 | −3.41 |
| K/SiNR-3 | −3.50 |

To elucidate the aforementioned phenomenon and determine the prefer adsorption sites and energies of the K atoms on the SiNRs, we conducted first-principles calculations. We chose the SNR and DNR as examples and calculated the adsorption energies of K atoms on three distinct adsorption sites. As illustrated in Figure 3, the sites are on top of the SNR (site 1), edge of the SNR (site 2), and the middle of the DNR (site 3). The computational results reveal distinct adsorption energetics for K atoms across different sites. At the edge site (site 2) of the SNR, the K adsorption energy reaches -3.41 eV, which is slightly more negative than the value of -3.40 eV observed at the central site (site 1). Notably, the most stable adsorption configuration occurs at site 3 on the DNR, where the adsorption energy reaches -3.50 eV, suggesting that this site facilitates the most stable K adsorption. These theoretical results agree with the STM topography results, confirming that site 3 is the preferred adsorption site for the K atoms. Additionally, various groups, including Sheng et al., have utilized noncontact atomic force microscopy to characterize the structures of SNRs and DNRs in real space, confirming the absence of chemical bonding between the two strands of the DNR using tip-enhanced Raman spectroscopy. In light of this information, we could reasonably speculate the adsorption of K atoms at multistranded SiNRs, for example, TPRs. The inset of Figure 2c shows the K atoms adsorbed at the TNRs in experiment, and Figure 4c illustrates the corresponding adsorption sites. Moreover, Figure 3e shows the schematic diagram of the zigzag structure formed during the saturation adsorption of K, corresponding to Figure 2d in experiment. These results provide significant insights into the adsorption behavior of K atoms on SiNRs.

It is worth highlighting that the K atoms adsorb on the SNR, DNR, and TNR display the same "triangular" feature. We attribute this phenomenon to two potential mechanisms. First, inspired by the formation of Co dimeric structures on SiNRs, we initially hypothesized that the "triangular" feature might arise from a trimeric K configuration. However, considering the atomic-scale interactions between K and its surrounding Si atoms, we ruled out this possibility due to the significant size mismatch between K and Si (with K having nearly twice the atomic radius of Si). Such a disparity makes it highly improbable for three K atoms to align precisely with the three underlying Si atoms to form a stable trimeric structure. In the second scenario, an isolated K atom would typically appear as a single protrusion in STM images. However, it is well established that STM signals reflect not only topographic height but also local density of states (LDOS). Detailed analysis of adsorption site 3 revealed that each K atom coordinates with three neighboring Si atoms. We propose that these Si atoms contribute significantly to the LDOS, which is subsequently detected by STM.

To validate this hypothesis, we calculated the charge density difference map of the SiNR following K adsorption, providing direct evidence for the electronic redistribution induced by the K-Si interaction. To confirm this scenario, we obtain the charge density difference map of the SiNR after K atom adsorption. In Figure 3d, the triangle marked by the dashed red line includes the morphological and electronic state information of the K atoms and their three nearest Si neighbors. The results reveal that a K atom is located at the center of a triangle formed by three Si atoms, and the K atom donates electrons; on the other hand, the electron density of the three closest Si atoms increases, resulting in electron accumulation around them. Consequently, both the K atoms and their three nearest Si neighbors contribute to the triangular feature observed in the STM images. This feature is similar to those previously reported for Na/Si(111)-(7 × 7) and K/silicene systems[12, 22, 37], where Na and K display distorted features owing to charge transfer to their nearest Si atoms. In addition, the charge density difference reveals that owing to the different positions occupied by the K atoms and their three nearest Si atoms, two distinct adsorption orientations are possible during charge transfer, which is consistent with our experimental results.

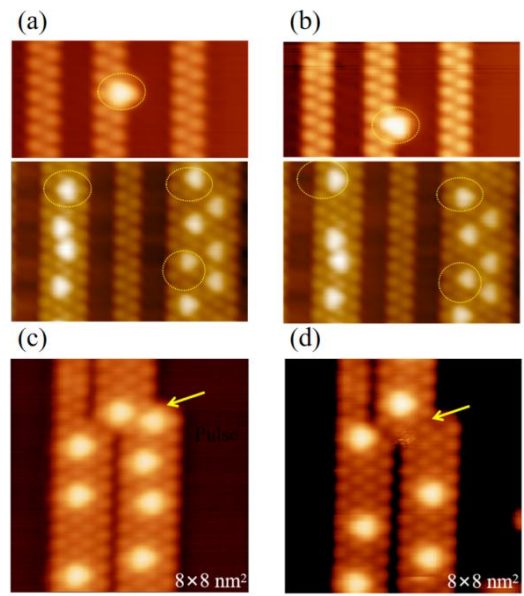

**Figure 4.** (a, b) STM images illustrating the variation in the orientation of K atoms adsorbed on SiNRs ($V = -800$ mV, $I = 100$ pA). (c, d) The STM images of K atoms adsorbed on DNRs before

and after the application of an appropriate tip pulse (5 V, 50 ms) (8×8 nm$^2$, $V = -800$ mV, $I = 100$ pA).

The stable adsorption site (site 3) and the triangular feature observed in the STM images are also supported by the hopping and diffusion of K atoms. According to previous reports, AM atoms usually have high mobility[22, 37]. During the experimental process, when the K atoms were repeatedly scanned, the triangular protrusion changes its orientation, accompanied by a small-scale movement of the K atom, as demonstrated in Figures 4a-4b. This movement may be due to the influence of the electric field induced by the tip or thermal interference during the scanning process. It indicates that the energy barrier to be overcome in this process is very low and even minor disturbances from the STM tip or other external factors are sufficient to cause the K atoms to switch to another adsorption orientation and move slightly, allowing the observation of this phenomenon in the STM topography. Regardless of how the K atoms moved, they remained adsorbed at site 3, and the triangle feature was observed for all adsorbed K atoms. Furthermore, we applied an appropriate pulse (5 V, 50 ms) *in situ*, directly above one K atom adsorbed to DNRs using the STM tip. A comparison of the STM topography before and after pulse application (Figures 4c-4d) reveals that the K atoms desorb from the SiNR after the pulse was applied, without damaging the SiNR structure. Notably, the SiNRs retain their original structure after the desorption of the K atoms by applying a pulse by the STM tip. Based on these experimental observations, we conclude that the adsorption of K atoms to SiNRs is a reversible adsorption process.

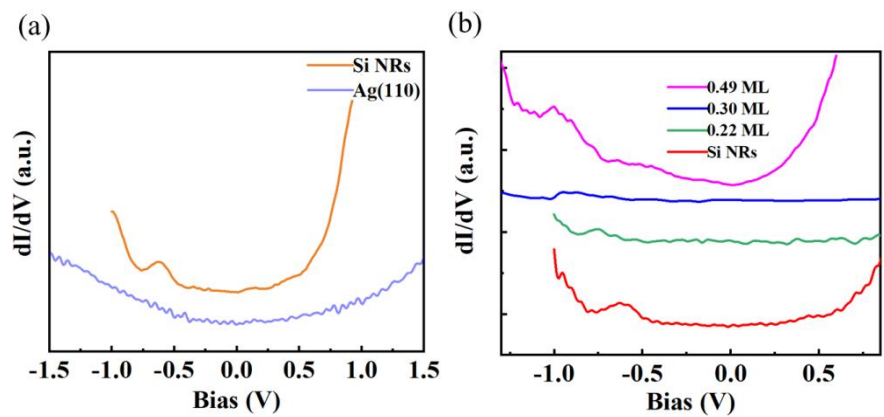

**Figure 5.** (a) Differential conductance spectra of the Ag(110) substrate and SiNRs at 77 K. (b) STS profiles of K atoms adsorbed on SiNRs at different coverages.

We performed STS measurements to further investigate the electronic properties of the SiNRs after the adsorption of K atoms. The STS spectra indicate that the bare Ag(110) substrate exhibits metallic behavior. While the SiNRs display a distinct dip feature at -800 mV, as illustrated in Figure 5a. This dip feature in the STS profile corresponds to the Dirac point of the SiNRs, consistent with the previously reported electronic information of the SiNRs based on STS and angle-resolved photoemission spectroscopy (ARPES)[19], thereby confirming the accuracy of our results for K adsorption on the SiNRs and the optimal conditions of the STM tip.

Next, we investigated the evolution of STS with increasing coverage of K. All STS spectra consistently exhibited a characteristic "dip" feature, which displayed a systematic downward energy shift of 279 meV below the Fermi level with the K coverage increasing. Specifically, at K coverages of 0.22, 0.30, and 0.49 ML, the "dip" feature was located at 0.866 V, 1 V, and 1.079 V below the Fermi level, respectively, as shown in Figure 5b. This result indicates that as the K coverage increases, more charge transfer occurs between the K atoms and SiNRs, leading to the movement of the Dirac cone below the Fermi level. Concurrently, the quantity of transferred charge carriers increases, and the K atoms form ionic bonds with Si. These observations are consistent with the theoretical framework proposed by Caragiu for the K-adsorbed graphite system, in which case the number of charge carriers transferred from K to the substrate is related to the downward shift of the Fermi level.[21]

Previous studies have elucidated the effects of K doping on the electronic structure of silicene-(3×3)[22, 38]. Notably, using ARPES, Feng et al. observed that at a critical level of K doping, the Dirac cone shifts by nearly 0.37 eV below the Fermi level[38]. Moreover, Qiu et al. determined that electron transfer from K to silicene-(3×3)/Ag(111) leads to a total shift in the binding energy of the Dirac cone by approximately 0.3 eV[22]. Our findings reveal that the Dirac cone of SiNRs exhibits a relatively smaller shift of 0.279 eV, which may be attributed to the lower coverage of K on the SiNRs compared with the close-packed adsorption of K on silicene. This result suggests that the extent of Dirac cone displacement is influenced by the density and arrangement of the K atoms on the Si-based material surface.

## 4. CONCLUSION

We utilized STM and STS to systematically investigate the adsorption behavior of K atoms on SiNRs and their resultant electronic properties. K atoms preferentially adsorb at interstitial sites between SiNRs, exhibiting enhanced thermodynamic stability. Each adsorbed K atom forms a distorted configuration on the SiNRs through charge transfer with nearest three Si atoms. Upon increasing the K coverage, these atoms self-assemble into 1D zigzag chains aligned along the SiNR. Remarkably, voltage-pulse manipulation via the STM tip enabled controlled switching of K atoms between two distinct adsorption orientations and small-scale displacements. Notably, high-amplitude pulses facilitated the complete desorption of K atoms without inducing structural damage to the SiNRs, demonstrating the fully reversible nature of the adsorption process. STS measurements further revealed that progressive K coverage induces significant charge transfer, as evidenced by a downward shift of the Dirac point below the Fermi level. This observation unambiguously confirms electron doping of the SiNRs and highlights the ionic character of the K-Si interactions. These results not only add new knowledge to the field of AM atom adsorption on 1D Dirac materials but also serves as an important reference for heteroatom adsorption on 1D NRs.

## AUTHOR INFORMATION

### Corresponding Author

**Wen-Xiao Wang** − Department of Physics and Hebei Advanced Thin Film Laboratory, Hebei Normal University, 050024 Shijiazhuang, China; Email: wangwx@hebtu.edu.cn

**Fengxian Ma** − College of Physics, Hebei Key Laboratory of Photophysics Research and Application, Hebei Normal University, 050024 Shijiazhuang, China; Email: fengxianma@hebtu.edu.cn

**Juntao Song** − Department of Physics and Hebei Advanced Thin Film Laboratory, Hebei Normal University, 050024 Shijiazhuang, China; Email: jtsong@hebtu.edu.cn

### Authors

**Tongtong Chen** − Department of Physics and Hebei Advanced Thin Film Laboratory, Hebei Normal University, 050024 Shijiazhuang, China

**Wenjia Zhang** − Department of Physics and Hebei Advanced Thin Film Laboratory, Hebei Normal University, 050024 Shijiazhuang, China

**Xiaobei Wan** − Department of Physics and Hebei Advanced Thin Film Laboratory, Hebei Normal University, 050024 Shijiazhuang, China

**Xiaohan Zhang** − Department of Physics and Hebei Advanced Thin Film Laboratory, Hebei Normal University, 050024 Shijiazhuang, China

**Yashi Yin** − Department of Physics and Hebei Advanced Thin Film Laboratory, Hebei Normal University, 050024 Shijiazhuang, China

**Jinghao Qin** − Department of Physics and Hebei Advanced Thin Film Laboratory, Hebei Normal University, 050024 Shijiazhuang, China

**Ying Liu** − College of Physics, Hebei Key Laboratory of Photophysics Research and Application, Hebei Normal University, 050024 Shijiazhuang, China; orcid.org/0000- 0002-7164-962X